\documentclass[twocolumn,showpacs,preprintnumbers,amsmath,amssymb]{revtex4-1}

\usepackage{afterpage}
\usepackage[dvipdfmx]{graphicx}
\usepackage{color,bm}

\begin{document}

\title{
Equivalence between definitions of 
the gravitational deflection angle of light for a stationary spacetime 
} 
\author{Kaisei Takahashi}
\email{kaisei@tap.st.hirosaki-u.ac.jp}
\author{Ryuya Kudo}
\email{kudo@tap.st.hirosaki-u.ac.jp}
\author{Keita Takizawa}
\email{takizawa@tap.st.hirosaki-u.ac.jp}
\author{Hideki Asada} 
\email{asada@hirosaki-u.ac.jp}
\affiliation{
Graduate School of Science and Technology, Hirosaki University,
Aomori 036-8561, Japan} 
\date{\today}

\begin{abstract} 
The Gibbons-Werner-Ono-Ishihara-Asada method for gravitational lensing in a stationary spacetime has been recently reexamined [Huang and Cao, arXiv:2306.04145], in which the gravitational deflection angle of light based on the Gauss-Bonnet theorem 
can be rewritten as a line integral of two functions $H$ and $T$. 
The present paper proves that 
the Huang-Cao line integral definition 
and the Ono-Ishihara-Asada one [Phys. Rev. D 96, 104037 (2017)] 
are equivalent to each other, 
whatever asymptotic regions are. 
A remark is also made 
 concerning the direction of a light ray in a practical use of these definitions. 
\end{abstract}


\maketitle

\section{Introduction}
The gravitational deflection of light plays a crucial role in modern cosmology 
and gravitational physics 
\cite{SEF,Petters,Dodelson,Keeton,Will}, 
where a conventional formulation of the gravitational deflection of light 
assumes the weak deflection of light in a quasi-Newtonian region 
that can be treated as a perturbation around Minkowski background. 

Although the conventional formulation is practically useful in 
many situations 
\cite{SEF,Petters,Dodelson,Keeton}, 
it is limited. 
In order to discuss a more geometrical aspect of 
the gravitational deflection of light, Gibbons and Werner (GW)
\cite{GW} 
proposed a use of the Gauss-Bonnet theorem (GBT) 
\cite{Carmo, Oprea}. 
The GW method was initially applied to 
a static and spherically symmetric (SSS) spacetime 
\cite{GW}, 
for which 
the deflection angle of light 
can be defined as a surface integral of 
the Gaussian curvature of the equatorial plane 
in the optical geometry. 
Later, Ishihara et al. generalized the GW idea 
for a case that an observer and source are located 
at a finite distance from a lens object 
\cite{Ishihara2016}. 
It was extended also to the strong deflection limit 
\cite{ Ishihara2017}. 
Without assuming the asymptotic flatness, eventually, 
Takizawa et al. 
proved the equivalence between the two definitions by 
GW and Ishihara et al. for SSS spacetimes 
\cite{Takizawa2020}. 

The GW method was extended by 
Werner to a stationary axisymmetric (SAS) case 
\cite{Werner2012}. 
This still employs asymptotically flat regions, 
at which the angle can be defined in a Euclid space. 
Furthermore, Ono, Ishihara and Asada (OIA) developed a formulation 
for a non-asymptotic observer and source 
in SAS spacetimes 
\cite{Ono2017}. 
These works assumed asymptotically flat regions. 
In the OIA approach, an alternative definition 
of the deflection angle of light 
was proposed in terms of a linear combination of three functions. 

It was proven \cite{Ono2017} 
that the deflection angle of light in the OIA approach 
is equivalent to the GW-type definition 
as a two-dimensional integral of the Gaussian curvature, 
if the SAS spacetime has asymptotically flat regions. 
See e.g. Eqs. (29) and (30) in 
\cite{Ono2017}.

Very recently, 
Huang and Cao (HC) have reexamined 
the Gibbons-Werner-Ono-Ishihara-Asada (GWOIA) method 
for SAS spacetimes 
\cite{HC}. 
They have found that 
the GW definition as a two-dimensional integral 
can be simplified as a line integral of two functions $H$ and $T$. 
See Eq. (44) in 
\cite{HC}. 

Can the OIA definition 
be related with the HC line-integral definition 
without assuming the asymptotic flatness? 
The main purpose of the present paper is to 
prove that the two definitions are equivalent to each other 
for SAS spacetimes, 
whatever asymptotic regions are. 

This paper is organized as follows. 
For its simplicity, 
first we consider a SSS spacetime 
to prove the equivalence in Section II. 
Section III extends the equivalence to SAS cases. 
Section IV summarizes this paper. 
Throughout this paper, we use the unit of $G=c=1$.

\section{Static and spherically symmetric case}
This section focuses on a SSS spacetime. 
The line element can be written as 
\cite{Ishihara2017}
\begin{align}
ds^2 = -A(r) dt^2 + B(r) dr^2 + C(r) (d\theta^2 + \sin^2\theta d\phi^2) .  
\label{metric-SSS}
\end{align}
In the rest of this section, 
we assume $A(r), B(r), C(r) > 0$. 
If the spacetime represents a black hole, 
we study the outside of a black hole horizon. 

Without loss of generality, 
a photon orbit can be chosen as the equatorial plane 
($\theta = \pi/2$) because of the spherical symmetry. 
From the null condition, we obtain \cite{Ishihara2017}
\begin{align}
dt^2 
&= \gamma_{ij} dx^i dx^j 
\notag\\
&= \frac{B(r)}{A(r)} dr^2 
+ \frac{C(r)}{A(r)} d\phi^2 ,
\label{dt2-SSS}
\end{align}
which defines the optical metric on the equatorial plane 
\cite{AK,GW,Ishihara2016}. 
We examine a light ray with an impact parameter $b$, 
which is related to the specific energy $E$ and 
specific angular momentum $L$ of a photon 
as $b \equiv L/E$. 
The null condition of a photon orbit becomes \cite{Ishihara2017}
\begin{align}
\left( \frac{dr}{d\phi} \right)^2 
+ \frac{C(r)}{B(r)} 
= 
\frac{C(r)^2}{b^2 A(r) B(r)} .
\label{orbiteq-SSS}
\end{align}
As a solution to Eq. (\ref{orbiteq-SSS}), 
$r$ along the light ray is a function of $\phi$.

\begin{figure}
\includegraphics[width=5.5cm]{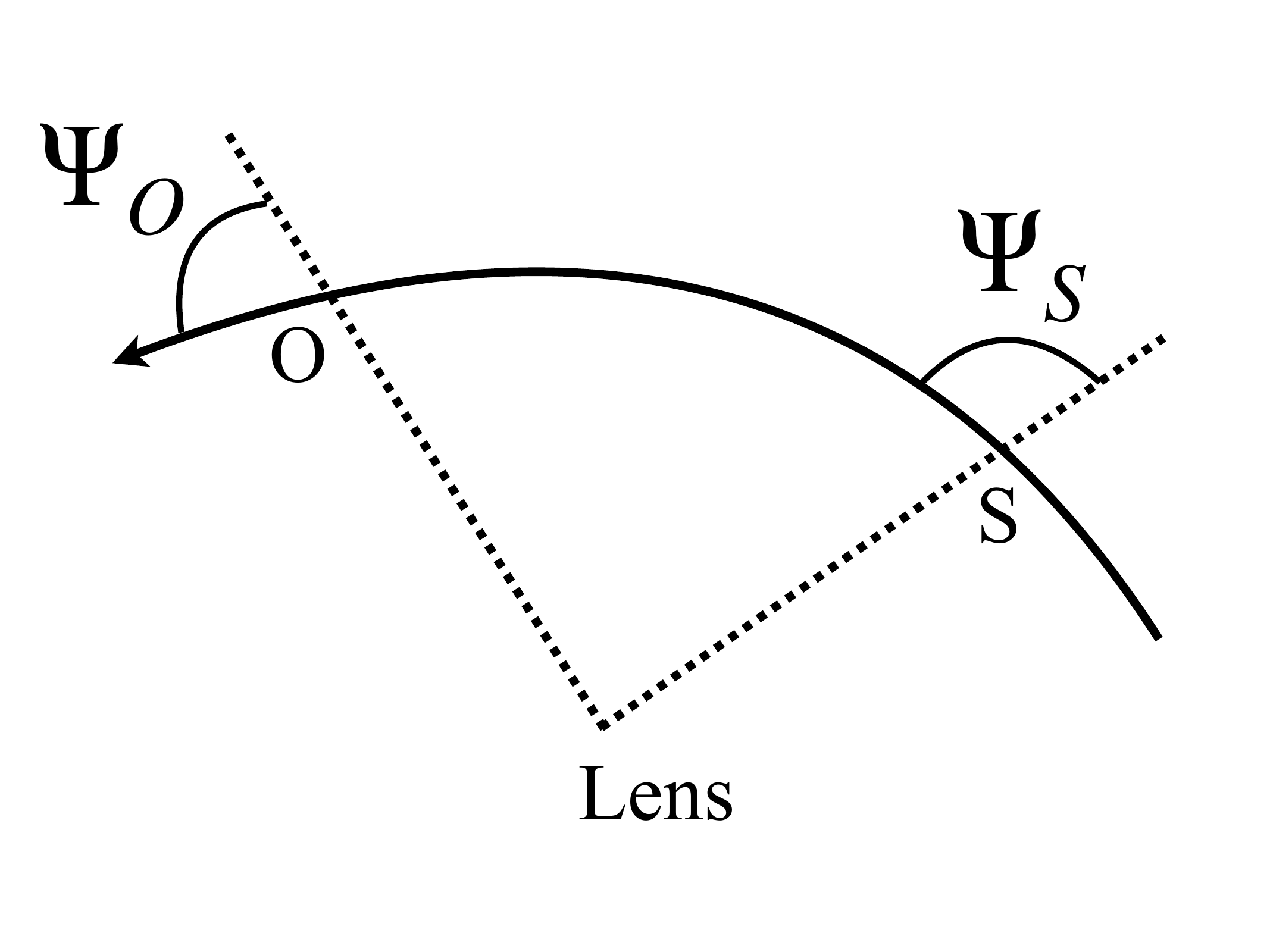}
\caption{
$\Psi$ as 
the angle from the radial direction to the light ray tangent. 
$\Psi_O$ and $\Psi_S$ are $\Psi$ at the observer and source, respectively.  
}
\label{fig-Psi}
\end{figure}

In the optical geometry, 
the angle from the radial direction to the light ray tangent 
is denoted as $\Psi$, 
which is expressed in terms of the metric components as 
\cite{Ishihara2016}
\begin{align}
\cos\Psi 
&= 
\frac{b\sqrt{A(r)B(r)}}{C(r)} \frac{dr}{d\phi} ,
\label{cosPsi-SSS}
\\
\sin\Psi
&= 
\frac{b\sqrt{A(r)}}{\sqrt{C(r)}} .
\label{sinPsi-SSS}
\end{align}
See also Figure \ref{fig-Psi}.

 By using $\Psi$ and $\phi$, 
Ishihara et al.  \cite{Ishihara2016,Ishihara2017} 
defines the deflection angle of light $\alpha_{I}$ as 
\begin{align}
\alpha_{I} 
\equiv \Psi_O - \Psi_S + \phi_{OS} ,  
\label{alpha-Ishihara}
\end{align}
where $\Psi_O$ and $\Psi_S$ are $\Psi$ at the observer (O) and source (S), 
respectively, 
$\phi_{OS} = \int_S^O d\phi$ is 
the longitude from S to O, 
and $\Psi_O$ equals to $\Psi_R$ in the notation of  
\cite{Ishihara2016,Ishihara2017}.
For the later convenience, 
this definition is rewritten as 
\begin{align}
\alpha_{I} 
= 
\int_S^O d\phi
\left( \frac{d\Psi}{d\phi} + 1 \right) .
\label{alpha-Ishihara-2}
\end{align}

By differentiating Eq. (\ref{sinPsi-SSS}) with respect to $\phi$, 
we obtain
\begin{align}
\frac{d\Psi}{d\phi}
= 
\frac{C(r)}{\sqrt{A(r)B(r)}} 
\frac{d}{dr}
\sqrt{\frac{A(r)}{C(r)}} ,
\label{dPsi-SSS}
\end{align}
where we use Eq. (\ref{cosPsi-SSS}) 
and $|dr/d\phi| < +\infty$ for a non-radial photon orbit.

See Eq. (4.25) 
in Reference \cite{HC} 
for the HC definition of the deflection angle of light. 
The HC definition is 
\begin{align}
\alpha_{HC} = 
\int^{O}_{S} 
d\phi [1 + H + T] .
\end{align}
For the SSS case, $H$ and $T$ simply become 
\begin{align}
H 
&\equiv 
- 
\frac{1}{2\sqrt{\gamma}} 
\frac{d (\gamma_{\phi\phi})}{dr} 
\notag\\
&= 
- \frac{A(r)}{2\sqrt{B(r)C(r)}} 
\frac{d}{dr} \left(\frac{C(r)}{A(r)} \right) ,
\label{H-SSS}
\\
T &= 0 ,
\end{align}
for $\gamma \equiv \det(\gamma_{ij})$.

The HC definition is thus reduced to 
\begin{align}
\alpha_{HC}
\equiv 
\int_S^O d\phi
\left( 1 + H \right) . 
\label{alpha-HC-SSS}
\end{align}

By direct calculations for Eqs. (\ref{dPsi-SSS}) and (\ref{H-SSS}), 
we find 
\begin{align}
H 
&=
\frac{C(r) A'(r) - C'(r) A(r)}{2 A(r) \sqrt{B(r) C(r)}} 
\notag\\
&= 
\frac{d\Psi}{d\phi} ,
\label{H-SSS-2}
\end{align}
where the prime denotes the differentiation with respect to $r$. 
Therefore, Eq. (\ref{alpha-Ishihara-2})  
equals to Eq. (\ref{alpha-HC-SSS}). 
In the SSS case, 
the two definitions are thus equivalent to each other.

\section{Stationary and axisymmetric case}
In this section, 
we consider a SAS spacetime. 
The line element can be written as \cite{Ono2017}
\begin{align}
ds^2 
=& 
-A(r, \theta) dt^2 + B(r, \theta) dr^2 + C(r, \theta) d\theta^2 + 
D(r, \theta)  d\phi^2 
\notag\\
&- 2 W(r, \theta) dt d\phi .  
\label{metric-SAS}
\end{align}

The null condition is rewritten in a form as 
\cite{Ono2017}
\begin{align}
dt = \sqrt{\gamma_{ij} dx^i dx^j} + \beta_i dx^i ,
\label{dt-SAS}
\end{align}
where  
\begin{align}
\gamma_{ij} dx^i dx^j
=&
\frac{B(r, \theta)}{A(r, \theta)} 
dr^2 
+
\frac{C(r, \theta)}{A(r, \theta)} 
d\theta^2
\notag\\
&+
\frac{A(r, \theta) D(r, \theta) + [W(r, \theta)]^2}
{[A(r, \theta)]^2}
d\phi^2 ,
\label{gamma-SAS}
\\
\beta_i dx^i 
=&
-\frac{W(r, \theta)}{A(r, \theta)} d\phi .
\end{align}

In the rest of this section, 
we focus on the equatorial plane ($\theta = \pi/2$) for a photon orbit, 
where we assume $A(r, \pi/2) , B(r, \pi/2), D(r, \pi/2) > 0$ 
and a local reflection symmetry with respect to $\theta = \pi/2$ 
as implicitly assumed in Reference 
\cite{Ono2017,HC}. 
Henceforth, 
$A(r, \pi/2), B(r, \pi/2), D(r, \pi/2) ,W(r, \pi/2)$ 
are denoted simply as $A, B, D, W$, respectively.

On the equatorial plane in the SAS spacetime, 
the OIA definition of the deflection angle of light is 
\cite{Ono2017}
\begin{align}
\alpha_{OIA} 
\equiv \Psi_O - \Psi_S + \phi_{OS} ,  
\label{alpha-Ishihara}
\end{align}
where $\Psi$ in the SAS metic satisfies 
\begin{align}
\cos\Psi 
&= 
\sqrt{\frac{B}{A}}
\frac{A (Ab + W)}{A D+ W^2} 
\frac{dr}{d\phi} ,
\label{cosPsi-SAS}
\\
\sin\Psi
&= 
\frac{Ab + W}{\sqrt{A D+ W^2}} .
\label{sinPsi-SAS}
\end{align}

By differentiating Eq. (\ref{sinPsi-SAS}) with respect to $\phi$, 
we obtain
\begin{align}
\frac{d\Psi}{d\phi}
= 
\sqrt{\frac{A D+ W^2}{AB}} 
\left(
\frac{(A b + W)'}{Ab + W}
-
\frac{(A D+ W^2)'}{2(A D+ W^2)}
\right) ,
\label{dPsi-SAS}
\end{align}
where we use Eq. (\ref{cosPsi-SAS}) 
and $|dr/d\phi| < +\infty$ for a non-radial orbit. 

The HC definition in the SAS case is 
\cite{HC}
\begin{align}
\alpha_{HC}
\equiv 
\int_S^O d\phi
\left( 1 + H + T \right) , 
\label{alpha-HC-SAS}
\end{align}
where $H$ and $T$ are defined as 
\begin{align}
H 
&\equiv 
- 
\frac{1}{2\sqrt{\gamma}} 
\frac{d (\gamma_{\phi\phi})}{dr} ,
\label{H-SAS}
\\
T
&\equiv
- \frac{d(\beta_{\phi})}{dr}
\sqrt{
\frac{1}{\gamma_{\phi\phi}}
\left(
\frac{dr}{d\phi}
\right)^2
+ 
\frac{1}{\gamma_{rr}} . 
\label{T-SAS}
}
\end{align}
In terms of the SAS metric components, 
$H$ and $T$ become 
\begin{align}
H
=& 
\sqrt{\frac{A D+ W^2}{AB}} 
\frac{A' D - D' A -2 W W' +2 W^2 A' A^{-1}}{2(A D+ W^2)} ,
\label{H-SAS-2}
\\
T
=&
\sqrt{\frac{A D+ W^2}{AB}} 
\frac{W' - W A' A^{-1}}{A b + W} .
\label{T-SAS-2}
\end{align}
By combining Eqs. (\ref{dPsi-SAS})
(\ref{H-SAS-2}) and (\ref{T-SAS-2}), 
one can show 
\begin{align}
H + T 
= 
\frac{d\Psi}{d\phi} .
\label{H+T}
\end{align}
Therefore, $\alpha_{OIA} = \alpha_{HC}$. 

Before closing this section, 
we mention the direction of a photon orbit.
The sign convention of $\Psi_O$, $\Psi_S$ and $\phi_{OS}$ 
in this paper is counterclockwise (See also Figure \ref{fig-Psi}). 
Hence, we should pay attention to the sign convention 
when we wish to distinguish prograde and retrograde motion. 
This issue seems a bit obscure in the HC line-integral definition, 
because $H$ and $T$ in Eqs. (\ref{H-SAS}) and (\ref{T-SAS})
are functions 
of the metric components and hence they do not directly manifest 
the direction of a photon (e.g. prograde or retrograde). 
The sign of  
$A b + W$ 
in $T$ of Eq. (\ref{T-SAS-2}) 
can distinguish prograde and retrograde.

\section{Summary}
We proved 
the equivalence between 
the OIA and HC definitions  
without assuming any property of the asymptotic regions 
in SAS cases, 
for which  
the GW-type definition also is equivalent to the HC one 
\cite{HC}. 
By combining the two results, 
the three definitions by GW, OIA and HC 
\cite{GW, Ono2017, HC} 
are equivalent to each other, 
whatever asymptotic regions are.

The essential part of the present proof 
relies upon the photon orbit 
but not upon any two-dimensional integration domain. 
This point agrees with the HC finding 
that the deflection angle in the Gauss-Bonnet method is 
independent of integration domains 
if the photon orbit is fixed 
\cite{HC}. 
The present proof thus deepens our understanding 
of the GBT-inspired definitions 
\cite{GW,Werner2012,Ishihara2016,Ishihara2017,Takizawa2020,Ono2017,HC}. 
Further study along this direction is left for future.

\begin{acknowledgments}
We are grateful to Marcus Werner for stimulating conversations. 
We thank Yuuiti Sendouda and Ryuichi Takahashi 
for the useful conversations.
This work was supported
in part by Japan
Science and Technology Agency (JST) SPRING, Grant
Number, JPMJSP2152 (K.T.), 
and JPMJSP2152 (R.K.)
and in part by Japan Society for the Promotion of Science (JSPS)
Grant-in-Aid for Scientific Research,
No. 20K03963 (H.A.).
\end{acknowledgments}

\end{document}